# Characterizing the Structure of 3D DNA Origami in a Transmission Electron Microscope

**Alyna Ong[1], Christoph Hadlich[2], Taekyu Jeong[3], Darius Pohl[1], Iman Elbalasy[2,4], Ralf Seidel[2], Michael Mertig[3, 5], and Bernd Rellinghaus[1].**

*[1] Dresden Center for Nanoanalysis (DCN), cfaed, TU Dresden, 01062 Dresden, Germany.*
*[2] Peter-Debye-Institute for Soft Matter Physics, Universität Leipzig, 04103 Leipzig, Germany.*
*[3] Physical Chemistry, TU Dresden, 01062 Dresden, Germany.*
*[4] Faculty of Science, Cairo University, 12613 Giza, Egypt*
*[5] Kurt-Schwabe-Institut für Mess- und Sensortechnik Meinsberg e.V., 04736 Waldheim, Germany.*

## ABSTRACT

DNA origami nanostructures provide programmable control over nanoscale geometry but remain challenging to image due to their low atomic number. Here, we systematically evaluate imaging strategies for both stained and unstained DNA origami deposited on carbon-coated TEM grids. Using Weber contrast as a quantitative metric, we compared different operating modes of (scanning) transmission electron microscopy, (S)TEM, in order to find optimum imaging conditions. STEM was consistently found to deliver the highest contrast, with optimal performance at a camera length of 600 mm towards the high angle annular dark field (HAADF) detector. As expected, the contrast was higher for the thicker three-dimensional nanotubes as compared to DNA 6-helix bundles (6HBs) due to the larger projected thickness of the former. The contrast was effectively enhanced by heavy metal staining with uranyl formate. Notably, 3D molds preserved their designated dimensions upon staining and also largely retained their structural integrity upon complexation with palladium, which also improved the visibility of the structures. These results establish STEM as the optimal approach for high-contrast imaging of DNA origami even of unstained samples and provide practical guidelines for sample preparation and imaging conditions that promote reliable structural visualization.

## INTRODUCTION

Emerging in the early 11080s, DNA nanotechnology leverages the programmability of Watson–Crick base pairing, the well-characterized B-form double helix, and the accessibility of custom sequence synthesis to enable the bottom-up construction of complex nanostructures [1-3]. A major advance occurred in 2006 with Rothermund's scaffolded DNA origami, in which hundreds of synthetic staple strands fold a long single-stranded DNA scaffold into prescribed shapes [1]. Through single-step thermal annealing, DNA origami reliably yields ~100 nm long nanostructures with efficiencies exceeding 90% [1,2,4,5], offering exceptional customization, spatial addressability, and nanometer-scale precision [6]. DNA origami can be engineered into diverse shapes and sizes for applications ranging from templating [7,8], molecular actuation [4,5,9], drug delivery [10], [11], biosensing [9-10], nanoelectronics [8] to molecular computing [8-10].

Precise structural characterization with high spatial resolution of such DNA origami is essential to validate designs [12], elucidate structural properties, and advance functionalities. Among the available tools, Atomic Force Microscopy, AFM, and (Scanning) Transmission Electron Microscopy, (S)TEM, are most widely employed [3,13]. AFM provides reliable imaging with (sub-)atomic height resolution but suffers from reduced lateral resolution [14-16] due to the unavoidable convolution of the sample with the tip geometry [3], while (S)TEM offers superior spatial resolution and compatibility with analytical techniques such as Energy Dispersive X-ray Spectroscopy, EDXS, and Electron Energy Loss Spectroscopy, EELS [3,17]. However, (S)TEM faces intrinsic challenges, specifically the low contrast obtained from materials that are predominantly composed of light elements – DNA origami is mainly composed of carbon, nitrogen and phosphorus – due to their weak electron scattering [5].

To overcome these limitations, two principal strategies are pursued. The first is the optimization of preparation techniques. This includes negative and positive staining, surface modification of (S)TEM grids through approaches like cation addition [3], plasma cleaning [3], or intercalation of electron dye [15,19], and the use of alternative carrier films such as ultra-thin amorphous carbon [3,13,19], amorphous silicon [13], graphene [17,19], and mica [3,13]. The second strategy aims at the electron optical imaging process, which includes methods such as high-angle annular dark field (HAADF) STEM [3], and in-focus phase-contrast TEM approaches like sub-Ångstrom low voltage electron microscopy, the use of volta phase plates, and dark-field microscopy [6].

Despite the significant advancements in (S)TEM methodologies, access to the latter techniques remains limited and depends frequently on the availability of according microscopes. Furthermore, while novel grid treatment strategies for sample preparation have demonstrated considerable promise, systematic investigations into the structural behavior or the folding states of DNA origami on surfaces remain scarce. In the present study, we provide a comparative analysis of conventional (in-focus) TEM, defocused TEM, and STEM applied to two types of dried DNA origami structures of different size (and scattering strength): thin 6-helix bundles (6HBs) [4,20] and larger hollow three-dimensional nanotubes further called molds [7,15]. Particular attention is paid to the optimization of the STEM parameters and the structural evaluation of the DNA origami under varying conditions, including negative staining, unstained preparations, and contrast enhancement by complexation with palladium complexes.

# RESULTS & DISCUSSION

## Imaging Modes and the Impact of Staining

Different imaging modes – including conventional in-focus transmission electron microscopy (TEM), defocused TEM, and scanning transmission electron microscopy (STEM) utilizing an annular dark-field detector – were evaluated for visualizing two types of DNA origami nanostructures: 6HBs and 3D molds with quadratic cross-section (see schematic illustrations in Figs. 1a,b). The samples were investigated in their pristine form and after being negatively stained with uranyl formate. The image contrast was assessed both qualitatively from mere visual inspection and quantitatively using Weber contrast [21]. This Weber contrast, $C_{Weber}$, is defined as the difference between the feature intensity and the background intensity, normalized by the background intensity. It was employed to identify the optimal imaging mode for a high-contrast visualization of these weakly scattering nanostructures:

$$C_{Weber} = \frac{I - I_{background\,(mean)}}{I_{background\,(mean)}} \qquad \textit{(Equation 1)}$$

Here $I$ denotes the image intensity as obtained from a line profile at the position of the object to be visualized, and $I_{background\ (mean)}$ is the mean intensity of the background as obtained from positions along the line profile that extend across the (stained) substrate only.

Each 6HB consists of a hexagonal arrangement of six parallel DNA helices each having an average length of 97.2 nm (represented by grey rods in the sketch in Figs. 1ab; for the calculation of the nominal length, c.f. Supporting Information Fig. S1). If stained with uranyl formate, the 6HBs appear as faint, hardly recognizable bright strings with dark shadows in images acquired by TEM and defocused TEM (cf. Figs. 1c and 1g), while they become visible as dark strings with bright halos in STEM images (Fig. 1k). The latter are due to the negative staining that leads of a preferred physisorption of the heavy metal-containing stain at the edges of the bundles. Without staining, the 6HBs remained basically invisible, when imaged using TEM, but they could be observed in STEM images, where they consistently appeared as bright strings (cf. Fig. 1l).

In order to quantify the contrast of these nanoscopic objects, line profiles of the Weber contrast are measured across the objects. Using these profiles, we define the feature contrast as the maximum change in Weber contrast upon crossing the object, i.e., $C_{Feat} = C_{Weber,max} - C_{Weber,min}$. Typically, these extrema occur at positions, where the change in the scattering strength (i.e., the mass-thickness of the objects) along the line of profile becomes maximal. In case of the 6HBs these are the central positions of the bundles themselves, while in case of the 3D molds, these are their side walls (see below). Such quantitative analysis confirms that STEM provides the highest feature contrast yielding $C_{Feat} = 0.32$ for stained and $C_{Feat} = 0.25$ for unstained 6HBs (Fig. 2a-b).

The 3D molds with nominal dimensions of approximately 40 nm × 20 nm × 20 nm showed a similar trend. If stained, TEM images revealed rectangular structures with two bright lines representing the two side walls of these hollow molds that adsorbed with the bottom side to the TEM grid (cf. Figs. 1e and 1i). In STEM images this contrast is inverted accordingly (cf. Fig. 1m). Unstained, the 3D molds can be imaged in alle three modes due to their higher thickness as compared to 6HBs. While in TEM images, the nonetheless weak contrast of the structures can be somewhat increased by defocusing (cf. Figs. 1f and 1j), STEM again provides for the best visibility (cf. Fig. 1n). Figs. 2c-d show as examples line profiles of the Weber contrast across one 3D mold each imaged with the different imaging modes. Here, the quantification yields feature contrasts of $C_{Feat} = 0.08$, $C_{Feat} = 0.25$, and $C_{Feat} = 0.32$ for negatively stained 3D molds imaged using in-focus TEM, defocused TEM, and STEM, respectively, while the corresponding feature contrasts in images of unstained 3D molds amount to $C_{Feat} \sim 0$, $C_{Feat} = 0.12$, and $C_{Feat} = 0.47$. Hence again, STEM clearly provides the best contrast for the imaging of the 3D molds.

From these observations, we deduce that STEM is the optimal imaging mode for visualizing DNA origami structures such as 6HBs and 3D molds both in their pristine form and after staining. This advantage arises from the way the electron beam interacts with the specimen and how scattered electrons are collected. In STEM, the imaging electrons are focused to form a small probe that is scanned across the specimen in a raster format [21-24], limiting the interaction with the sample to the position of the focused beam. Since the scattering strength of the high energy electrons scales with $Z^{1.5} - Z^{2}$ with $Z$ being the (effective) atomic number [25], thickness variations can be very effectively monitored on the HAADF detector using STEM. In contrast, in TEM, the interference of coherent partial beams of the incoming planar electron wave results in a (close to) vanishing phase contrast on top of the large signal of unscattered electrons in light element (weak phase object) materials [21]. By collecting only scattered electrons (with even some *Z*-specificity) and excluding unscattered electrons, the HAADF detector improves the signal-to-noise ratio and provides superior contrast relative to TEM. Furthermore, the effects of chromatic lens aberration are less relevant in STEM [21,23,24,26], where the resolution is mainly determined by the probe size.

While STEM has been shown to be the optimal imaging mode, defocused TEM still provides for an improved contrast compared to conventional in-focus TEM. This improvement is due to the aberrations that are deliberately inferred from defocusing the sample. The latter introduces an additional phase shift to the interfering electron waves that results in an enhanced, though non-uniform (spatial frequency-dependent) and partially de-localized image contrast. Accordingly, defocused TEM provides for better visibility of the DNA origami than in-focus TEM.

When comparing the unstained DNA origami samples (Figs. 2b,d), 3D molds exhibit a higher contrast than 6HBs across all imaging modes. This difference arises from the higher mass-thickness contrast of the latter. The larger thickness of the side walls of the 3D molds along the beam direction (~ 20 nm) compared to 6HBs (~ 4 nm in the center of the bundle) increases the number of elastic scattering events. Consequently, more electrons are collected on the HAADF detector.

Staining also contributes to an improved visibility [26]. Uranyl formate, a high-atomic number material, enhances the electron scattering, resulting in an increased contrast in both TEM and STEM. In in-focus and defocused TEM (Fig. 1), uranyl formate deposits appear as dark regions, since where ever it is residing, less electrons are transmitted in the forward direction through the sample. In STEM, the same deposits rather appear bright, as the strongly scattered electrons are now collected by the HAADF detector. However, there is also a trade-off of the staining: Negatively stained 3D molds exhibit a significantly weaker feature contrast of $C_{Feat} = 0.32$ than unstained 3D molds ($C_{Feat} = 0.47$) (Fig. 2c-d). This reduction is due to residual uranyl formate that is also deposited in the local environment of the objects to be stained, which leads to an enhanced background intensity and thereby a reduced overall contrast.

In summary, STEM provides for the highest feature contrast of both stained and unstained DNA origami, thereby outperforming in-focus TEM and defocused TEM imaging. These findings establish STEM as the optimal imaging mode for DNA origami and highlight key physical factors that determine visibility, offering practical guidelines for future structural and quantitative studies of weakly scattering biomolecular assemblies.

## Optimization of STEM imaging by adjusting the camera length

Based on this result, we optimized the parameters for STEM imaging in a second step. This is essential, because the scattering angle of the electrons varies with the atomic number of the scatterers *Z* and thus with the material composition [21,24]. In STEM mode, the angular range of electrons arriving at the HAADF detector is controlled by adjusting the camera length *L* that is defined as the effective distance from the back focal plane of the microscope to the detector. Adjusting the projector lens excitation alters the magnification of the diffraction pattern, thereby effectively modifying the camera length.

To identify the optimal STEM imaging conditions for unstained 3D molds, we tested four different camera lengths: 400 mm, 600 mm, 800 mm, and 1000 mm. The resulting STEM images obtained from identical sample positions are shown in Figs. 3b–e. Line-profiles of the image intensity across the same 3D mold were then acquired, from which profiles of the Weber contrast were calculated. As shown in Fig. 2a, the contrast is the highest for $L$ = 600 mm. Since both smaller and higher camera lengths resulted in a comparable lower contrast, a camera length of 600 mm proved to be the optimum for imaging DNA origami structures. This parameter is thus used in subsequent experiments and offers a reference for future quantitative studies of DNA origami nanostructures using STEM.

## The impact of staining on the morphology of 3D molds

Stained 3D molds are seen as rectangular features bordered by two pronounced dark lines surrounded by bright halos along their long edges in the STEM micrograph (Fig. 4a, see also above). The apparent contrast arises from the negative staining with uranyl formate that physisorbs at the outer edges of the DNA origami. As a consequence of the higher scattering strength of the heavy metal ions in the stain, the outer rims of the molds appear brighter than the encapsulated DNA origami. Statistical analysis of many of such images reveal that the mean projected dimensions (42.4 ± 1.3) nm × (23.0 ± 1.6) nm (cf. Fig. 4d) closely match the theoretical size of 20 nm × 40 nm. This shows that the uranyl formate stain preserves the intrinsic three-dimensional architecture of the molds.

In contrast, without staining, the molds experience a lateral expansion of their width (Fig. 4b). On average the observed rectangular structures have projected dimensions of (41.2 ± 1.8) nm × (29.8 ± 2.0) nm. Thus, the length of the molds remains largely unaltered with respect to the expected value, while their (projected) width is clearly enhanced. This indicates that upon deposition on the amorphous carbon carrier film, the 3D molds are morphologically instable and collapse (partially) into flattened structures. While the chemical bonds between the individual DNA helices seem to prevent a shearing of the structure along the direction of the helical axes, a lateral collapse that would only necessitate minor mutual rotations of neighboring helices appears feasible.

In comparison, Fig. 4c shows as an example the STEM image of 3D molds that were "stained" by complexation with $[PdCl_4]^{2-}$ complexes (further referred to as Pd molds) prior to the deposition on the carbon film. The binding of multivalent $[PdCl_4]^{2-}$ complexes is known to provide an extraordinary thermal stabilization of DNA nanostructures [15]. In the images one observes two parallel lines that are brighter than the cavity. Due to the Pd enrichment in the walls of the hollow molds, the upright side walls are aligned parallel to the imaging electron beam providing increased contrast, while the contrast of the perpendicular bottom and top side walls is reduced. This clearly indicates that the Pd complexation helps to preserve the three-dimensional structure of the molds and prevents them from collapsing, although less Pd complexes than DNA nucleotides are needed for the stabilization. This confirms that Pd-complexed DNA nanostructures exhibit an extraordinary stability even in the dried state. Noteworthy, the statistical evaluation of the projected geometries of the Pd molds reveals a slightly reduced length of (37.9 ± 2.6) nm as compared to the negatively stained ones but a similarly enhanced width of (28.8 ± 6) nm. Both findings can be attributed to the kinking of the DNA helices upon multivalent binging of the $[PdCl_4]^{2-}$ complexes [15], the statistical nature of which also explains the significant broadening of the size distributions as compared to those of the negatively stained molds (cf. Fig. 4e), which results in an enhancement of the width of the Pd molds by upto 50%.

## Morphology of unstained 3D molds in the dry state

The apparent collapse of the unstained 3D molds in their dry state and after deposition onto the carbon carrier films as deduced from the quantitative determination of their projected geometry is further supported by a semi-quantitative analysis of the intensity profiles across such structures obtained from STEM images. To this end we compared the profiles of the normalized HAADF intensities across a single unstained 3D mold for all investigated camera lengths (Fig. 5). Interestingly the normalized profiles were indistinguishable suggesting a high accuracy and reliability of the geometrical measurements above.

Furthermore, this observation provides evidence that the HAADF intensities can be directly related to the mass-thickness of the sample and thus the number of scatterers in the direction of the electron beam. A pristine 3D mold in solution is composed of 64 helical DNA strands with approximated diameters of 2 nm each (see upper sketch in Fig. 5). Hence, in their uncollapsed, pristine state, the mass-thickness profile of the mold should have a width of some 20 nm with pronounced maxima at its outer rims. This is in stark contrast to the experimentally determined profile that is much wider and rather exhibits a bell-shaped form. Assuming that upon a morphological collapse, the top and bottom horizontal walls of the formerly hollow structure come to lie on top of each other, the maximum HAADF intensity in the middle of the profile would correspond to four layers of DNA helices stapled in the direction of the imaging electron beam. Translating under this assumption the measured intensity profile to a profile of the number of stapled DNA strands provides also a bell-shaped height profile (Fig. 5, left sketch). Such a profile may be obtained by a collapse of two pairs of neighboring side-walls onto each other, i.e. across the diagonal of the quadratic cavity (Fig. 5, right sketch). Though the exact details of the final structure remain unresolved, our data evidently support the collapse of molds in their dry, deposited state. Together with the above geometrical analysis, this implies that the gain in interfacial energy upon depositing the 3D molds on the carbon substrate is larger than the chemical bond-related energetic stabilization of their hollow morphology in liquid (or vacuum).

The observed collapse of the 3D molds is consistent with structural distortions previously reported for other 3D DNA origami imaged on solid supports by either AFM [14-16] or TEM [27]. AFM studies frequently result in reduced heights of DNA origami relative to their designed dimensions [14,15]. Specifically, the findings of Kemper et al. support our conclusions [15]: In their study, the authors have used AFM to measure the height of 3D mold and report that pristine 3D molds extend only to roughly half their theoretical height, if they are deposited on a substrate. In contrast, Pd-complexed molds exhibit a height of 25 nm in agreement with their measured widths. Also tetrahedral DNA nanostructures (TDNs) were observed to completely collapse upon deposition on bare highly oriented pyrolytic graphite (HOPG) due to strong substrate interactions despite their rigid 3D design [14].

## Quantitative analysis of 6HB DNA origami

Finally, the impact of negative staining on the lengths of 6HBs was investigated. In the STEM micrographs, the 6HBs appear as rod- or string-like features (cf. Figs. 1k and 1l). Fig. 6 shows the histograms of lengths determined from such STEM images of both negatively stained and unstained bundles. Again, the size distributions were fitted by Gaussian distribution functions, from which the resulting mean projected lengths of the stained and unstained 6HBs were determined to be (95.7 ± 5.8) nm and (91.2 ± 8.3) nm, respectively. While experimentally determined lengths of the stained 6HBs are in good agreement with the 6HBs' mean theoretical length of 96.4 nm (cf. Supporting Information), the unstained 6HBs are found to be slightly shorter than expected from the mere length and arrangement of the DNA helices they are composed of. Technical aspects may certainly limit the accuracy of length measurements on projection images. For example, the length of slightly bent structures may be falsely determined depending on the type of (software) measurement tool used. In addition, fuzzy boundaries arising from staining residues or single-stranded DNA overhangs may obscure structural terminations and this way result in an underestimation of their length [28]. Also, dehydration of the samples during preparation or sample – substrate interactions may alter the DNA origami conformations, thereby promoting their compaction after deposition [15].

# CONCLUSIONS

In conclusion, our study demonstrates that STEM is the most effective imaging mode for visualizing DNA origami structures, owing to its superior signal-to-noise ratio and contrast when compared to TEM. While TEM benefits from phase-contrast effects introduced through defocusing, STEM's use of a focused probe and HAADF detection enables precise monitoring of mass-thickness variations, particularly for unstained samples. Our comparative analysis revealed that 3D molds consistently exhibit a higher contrast than 6HBs across all imaging modes due to their higher mass-thickness. Staining with uranyl formate further enhances visibility due to the enhanced scattering of its heavy metal ingredients, though this comes at the cost of increased background intensity and a slightly reduced feature contrast. Importantly, we identified a camera length of 600 mm as the optimal STEM parameter for imaging DNA origami.

Beyond this optimization of the microscopic imaging, our results provide insights into the structural behaviour of different DNA origami architectures upon depositing them in a dried state on a substrate. For 3D molds, uranyl formate staining preserved their intrinsic three-dimensional geometry, whereas unstained samples showed lateral expansion due to strongly enhanced adhesion forces as a consequence of the drying on the substrate. Pd-complexation offers an alternative stabilization mechanism preventing the structural collapse of the Pd molds and maintaining their structural integrity with fewer Pd complexes than nucleotides, even under dehydrating conditions. A significant broadening of the Pd molds as compared to the negatively stained ones is ascribed to the kinking of the helices that goes along with the Pd-complexation. Similarly, also 6HBs could be stabilized to their theoretical length by negative staining, while the lengths of unstained bundles were slightly reduced.

In summary, our findings underscore that both imaging mode and sample preparation significantly influence the morphology of these nanostructures, and that careful consideration is required when interpreting structural features of DNA origami on the basis of microscopic images. This work therefore lays the foundation for more accurate structural investigations, which are crucial for advancing the design and application of DNA nanotechnology.

# METHODOLOGY

## DNA origami synthesis

Six-helix-bundles (6HBs) were synthesized by folding from truncated single-stranded M13mp18 virus scaffold DNA (1,702 nucleotides long) with 40 mM Tris base, 20 mM $MgCl_2$, 2 mM EDTA buffer as previously described [4,5,29].

For the assembly of the 3D molds a 10 nM of long single-stranded scaffold strand type p8064 (Eurofins), based as well on the M13mp18 virus, was mixed with the 10-fold concentration of staple strands (Eurofins)and 5 mM Tris-HCl, 1 mM EDTA, 5 mM NaCl and 11 mM $MgCl_2$, pH 8.0 [7]. In a one-pot reaction the solution was first heated to 80 °C for 5 min and then cooled to 25 °C over 15 h using a nonlinear temperature ramp [7]. Afterwards, the assembled structures were purified from excess staples by polyethylene glycol precipitation [30]. Pd-based stabilization of the molds was achieved by employing the recipe provided by U. Kemper et al. [15]. For this, an aliquot of 10 mM $[PdCl_4]^{-2}$ stock solution was diluted to 1 mM directly before mixing it with the origami solution containing 10 nM of 3D molds and 500 mM boric acid as buffer. Final concentrations were 6 nM of DNA origami structures and 24 µM $[PdCl_4]^{-2}$. Structures were incubated overnight and then dialyzed against 2 L of boric buffer solution using a 2 kDa cut-off dialysis unit (Slide-A-Lyzer® MINI, Thermo Fisher Scientific) [15].

### Staining solution preparation

In a beaker, 5mL of 100 °C deionized water were added to 100 mg of uranyl formate. The solution was stirred for 5 mins, filtered with a 0.22 µm filter and distributed into several 100µL aliquot. The aliquots were wrapped in aluminum foil and stored at -20°C. To use for staining, 2.5 µL of 1M sodium hydroxide was added and the 100 µL aliquot shook for 30 mins. at room temperature (22 °C).

### Deposition of DNA origami on a TEM grid

TEM grids (Carbon Square Mesh, Cu, 200 Mesh, UL) purchased from Electron Microscopy Sciences were plasma cleaned with 95% argon gas and 5% oxygen gas in FISCHIONE Model 1020 Plasma Cleaner, for 1 min, at 10.55 W and 30 sccm gas flow. Unlike otherwise mentioned, all TEM grids for all sample were plasma cleaned.

To prepare unstained DNA origami samples: On plasma cleaned TEM grids, a 3µL droplet of DNA origami solution was incubated for 5 mins. The liquid was removed by tapping perpendicular to a filter paper and the grids were air dried for at least 30 mins.

To prepare stained DNA origami samples: On plasma cleaned TEM grids, a 3µL droplet of DNA origami solution was incubated for 5 mins. The liquid removed by tapping perpendicular to a filter paper. A 3 µL droplet of uranyl formate solution was immediately dropped and removed. Another 3 µL droplet of uranyl formate solution was dropped, incubated for 10s and removed. TEM grids were air-dried for at least 30 mins. Samples containing stained 3D molds were cleaned again at the end with 5 µL of deionized water.

### (S)TEM imaging

(S)TEM images were acquired using a JEOL JEM-F200 multiple-purpose transmission electron microscope, equipped with a cold FEF and a GATAN OneView CMOS camera operating at 200kV.

### Measurement, image analysis and processing

Images were analyzed using the GATAN DIGITAL MICROGRAPH software. All micrographs were adjusted equally for brightness and contrast to enhance feature visibility and to standardize the background appearance. No other image processing was performed. The in-built functions “curve” and “straight line” were used to measure the length of the 6HB and both (long and short) axes of the 3D molds, respectively. Gaussian fits (with orthogonal distance regression and instrumental weighting of a 2% error in the measured sizes) to the distributions of likewise measured lengths were used to determine mean measures of the DNA origami. The centers and half widths of these Gaussian fits were then used as the mean values and uncertainties of the according size distributions, respectively.

# AUTHOR CONTRIBUTIONS

A.O. conducted the TEM experiments, processed the images, and analyzed the data with the help of B.R., M.M, R.S., and R.S. D.R. C.H., T.J., and I.E. prepared the samples. B.R., M.M., and R.S. devised the experiments and supervised the work. All authors contributed to the discussion of the results. A.O. and B.R. wrote the manuscript with the help of R.S. All authors contributed to the discussion of the results.

# ACKNOWLEDGEMENTS

Funding by the Deutsche Forschungsgemeinschaft (DFG, German Research Foundation) within GRK 2767 (project no. 451785257) is gratefully acknowledged.

# CONFLICTS OF INTEREST

There are no conflicts of interest to declare.

# DATA AVAILABILTY

There are no conflicts of interest to declare. The data are not publicly available. However, the data are available from the authors upon reasonable request.

# FIGURES

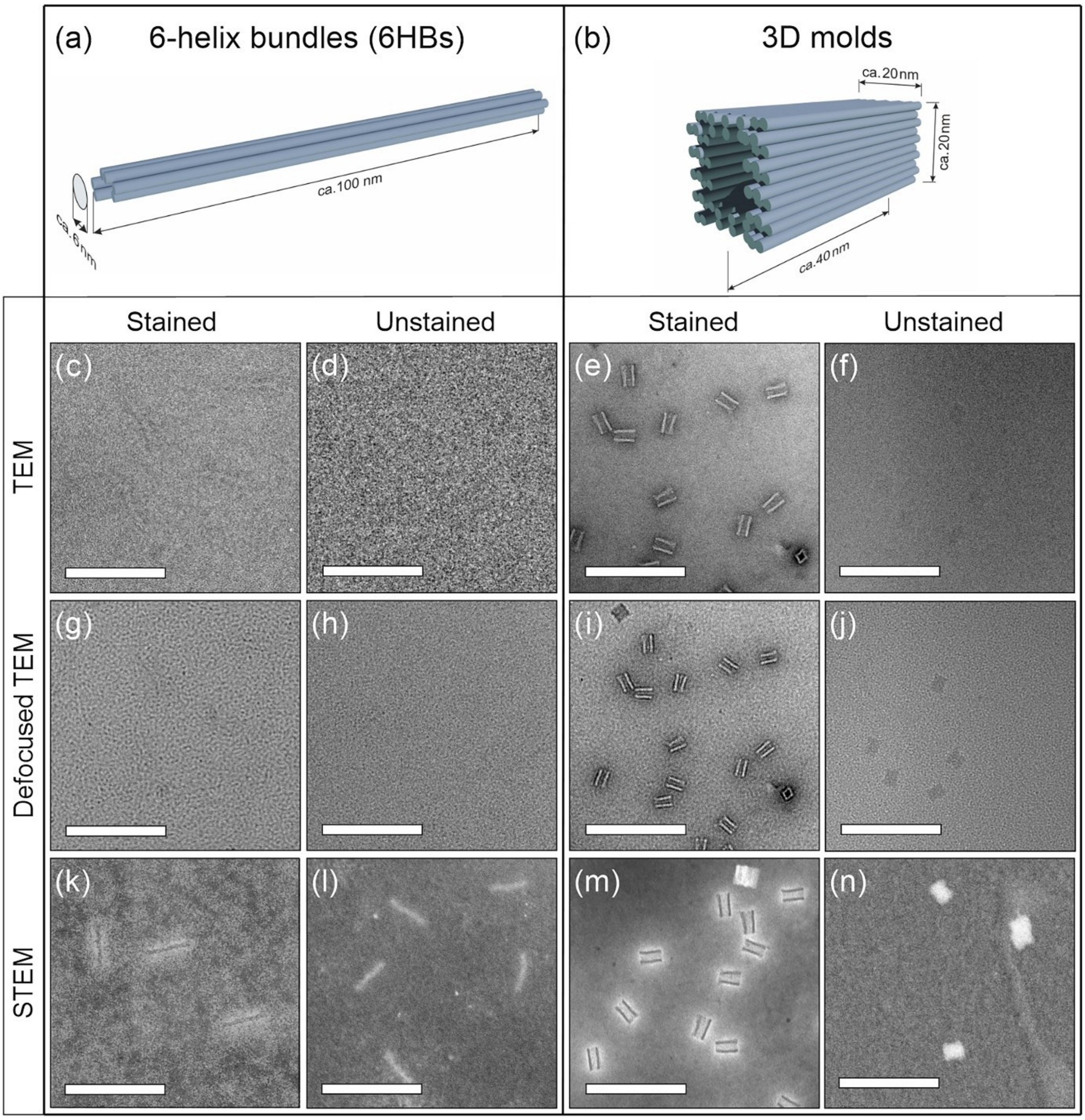

***Figure 1****: Transmission Electron Microscopy (TEM) of DNA origami: Images of (a) 6HBs (left columns) and (b) 3D molds (right columns) are acquired using (c-f) conventional in-focus TEM, (g-j) defocused TEM, and (k-n) STEM. Both, negatively-stained and unstained samples are characterized. Stained with uranyl formate, both 6HBs (albeit partially faintly) and 3D molds can be visualized in all imaging modes, with a string-like appearance of the 6HBs and a clear visualization of the side walls of the hollow 3D molds. Without staining, 6HBs were detectable only in STEM mode, whereas 3D molds remained observable in both defocused TEM and STEM modes. Scale bar is 200 nm. The contrast of all images is adjusted for uniform illustration.*

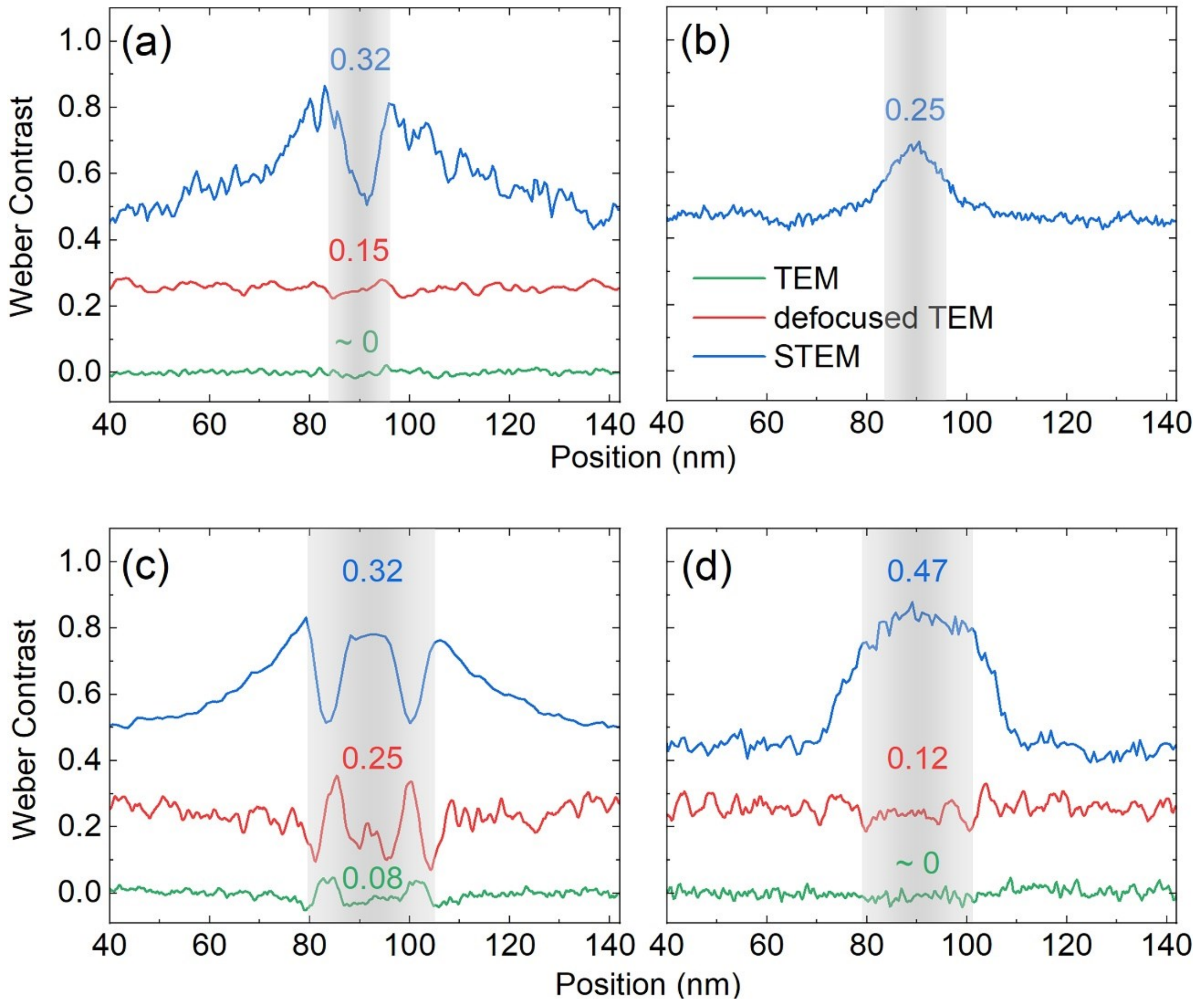

***Figure 2:*** *Line profiles of the Weber contrast across individual, differently prepared 6HBs and 3D molds: (a) 6HB negatively stained with uranyl formate, (b) unstained 6HB, (c) negatively stained 3D mold, and (d) unstained 3D mold. Each panel includes profiles obtained from images acquired by conventional in-focus TEM (green lines), defocused TEM (red), and STEM (blue). Grey shaded bars are guides to the eye and indicate the mean widths of the according DNA origami. For clarity, the profiles are mutually offset. Coloured numbers denote the resulting feature contrast for each individual profile. In all cases, STEM (blue lines) provides for the highest contrast.*

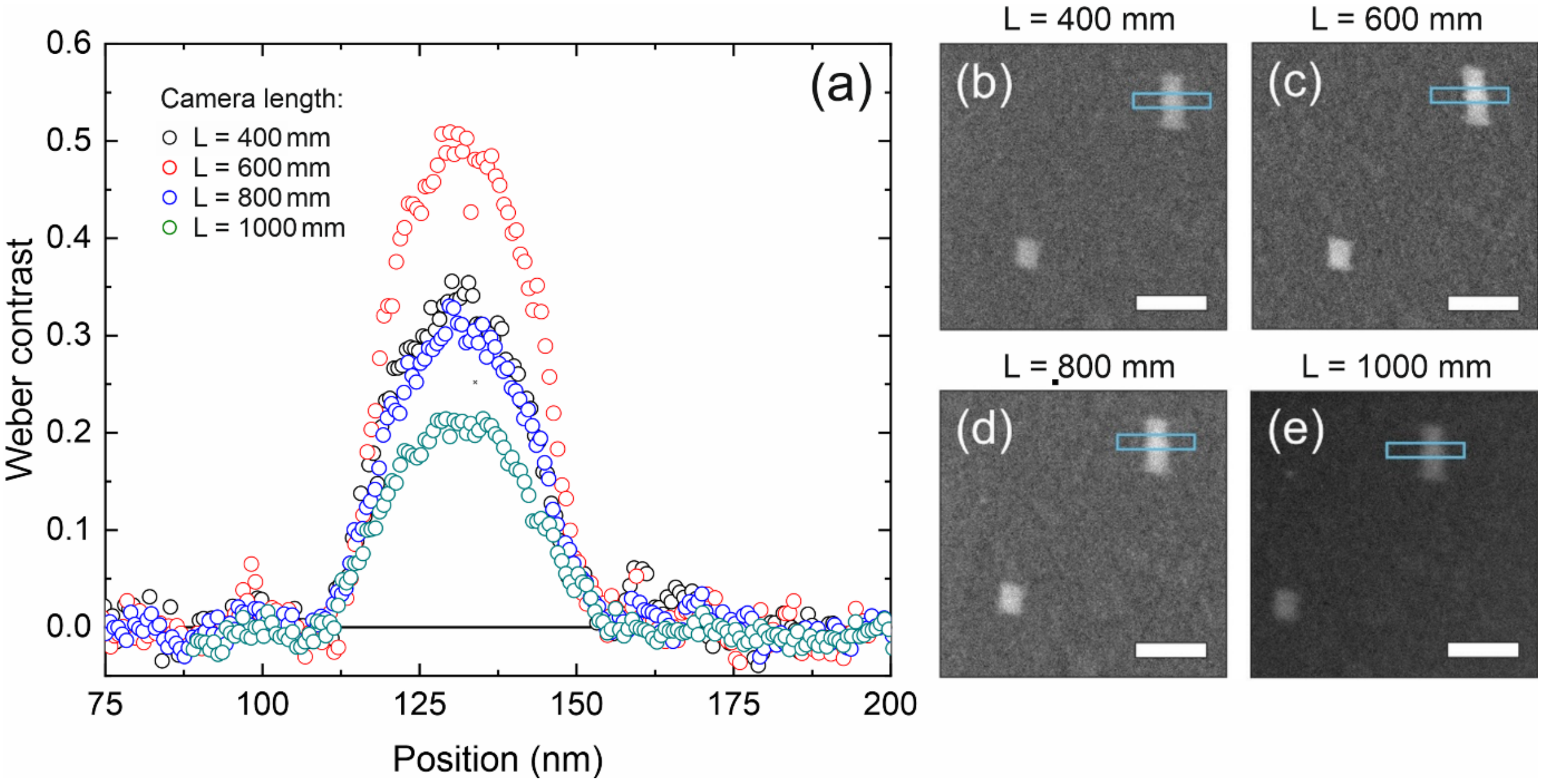

***Figure 3****: Effect of the camera length on the contrast of 3D molds in STEM images of unstained 3D molds. Images were acquired at camera lengths, L, of 400, 600, 800, 1000 mm. (a) Weber contrast profiles (centered to identical positions) across an identical 3D mold reveal that a camera length of 600 mm provides for the highest contrast between the unstained 3D molds and the background (which compares to the feature contrast): The determination of the feature contrast from these profiles yields* $C_{Feat}$ *= 0,34, 0.49, 0.32, and 0.21 for camera lengths of 400, 600, 800, and 1000 mm, respectively. (b-e) Corresponding STEM images. Scale bars are 100 nm each.*

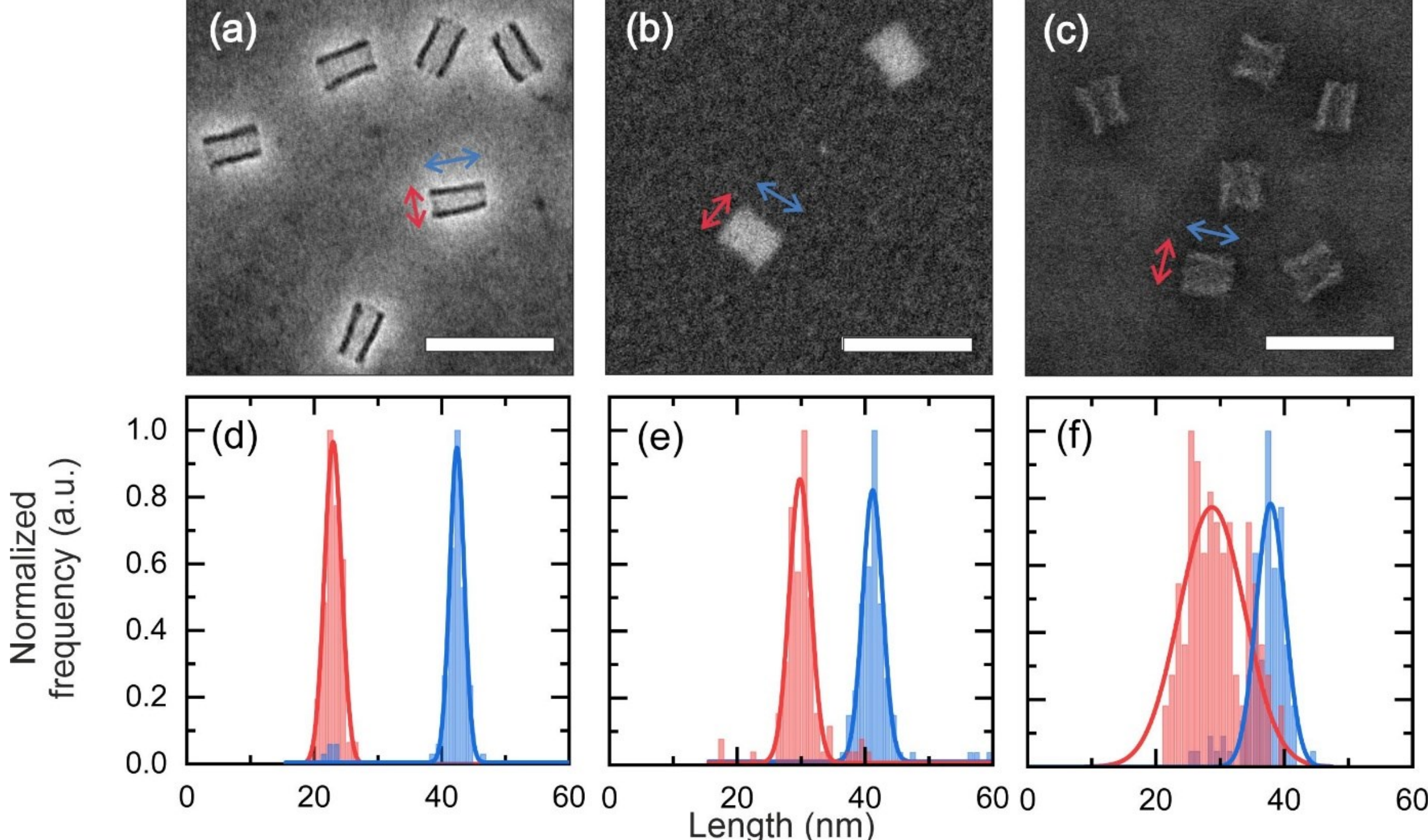

***Figure 4****: Examples of STEM images of 3D molds prepared under different conditions: (a) negatively stained with uranyl formate, (b) unstained on the as-received carbon support film of a commercial TEM grid, and (c) intercalated with Pd. Scale bars are 100 nm. (d-e) Statistical analyses of the geometrical measures of the likewise prepared 3D molds: Histograms of the lengths (blue) and widths (red) of the 3D molds. Resulting dimensions are (d) 42.4 ± 1.3 nm and 23.0 ± 1.6 nm for the negatively stained molds, (e) 41.2 ± 1.8 nm and 29.8 ± 2.0 nm for the unstained ones, and (f) 37.9 ± 2.6 nm and 28.8 ± 6.0 nm for the Pd-intercalated molds. These values represent the centers ± the half widths of the distributions as determined from Gaussian fits to the experimental data.*

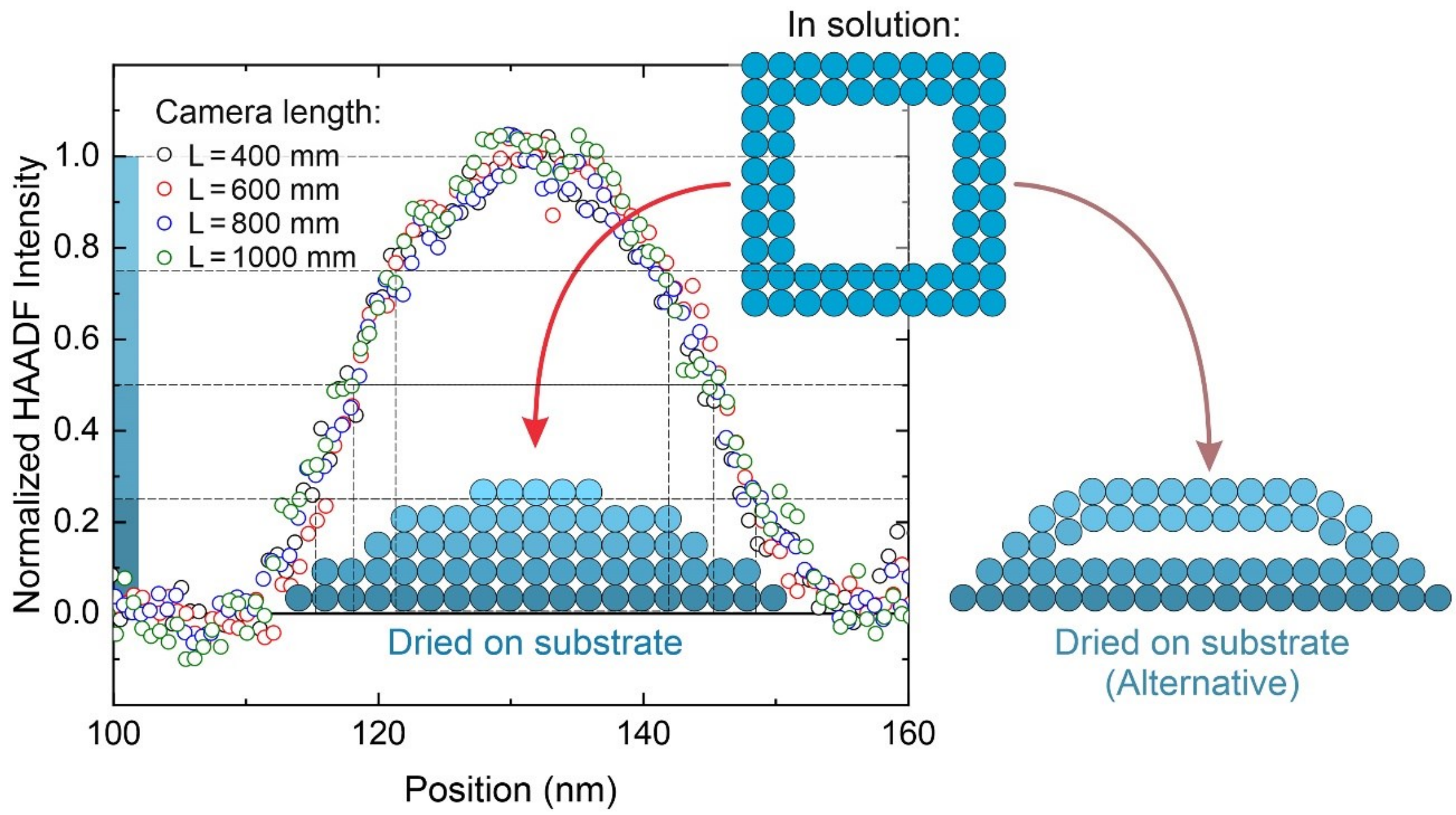

***Figure 5****: Semi-quantitative analysis of the intensity profiles across an unstained 3D mold as derived from STEM images acquired at different camera lengths. The background-subtracted HAADF intensities are normalized to their maxima, and the profiles are aligned to a common center position. As the intensities in a STEM images are due to the mass-thickness of the sample, the measured intensities are proportional to the number of scatterers in STEM images. Hence, in a simplistic approach, the number of individual DNA strands along the beam direction (with diameters of roughly 2 nm each – represented by blue circles drawn to scale of the abscissa) can be deduced from these profiles (see text for details). Due to the nature of this projection, the distribution of the strands along the beam direction, i.e., their height distribution, cannot be inferred from these profiles. As a consequence, and as indicated by the alternative sketches, different undistinguishable 3D arrangements of the apparently collapsed 3D molds are possible.*

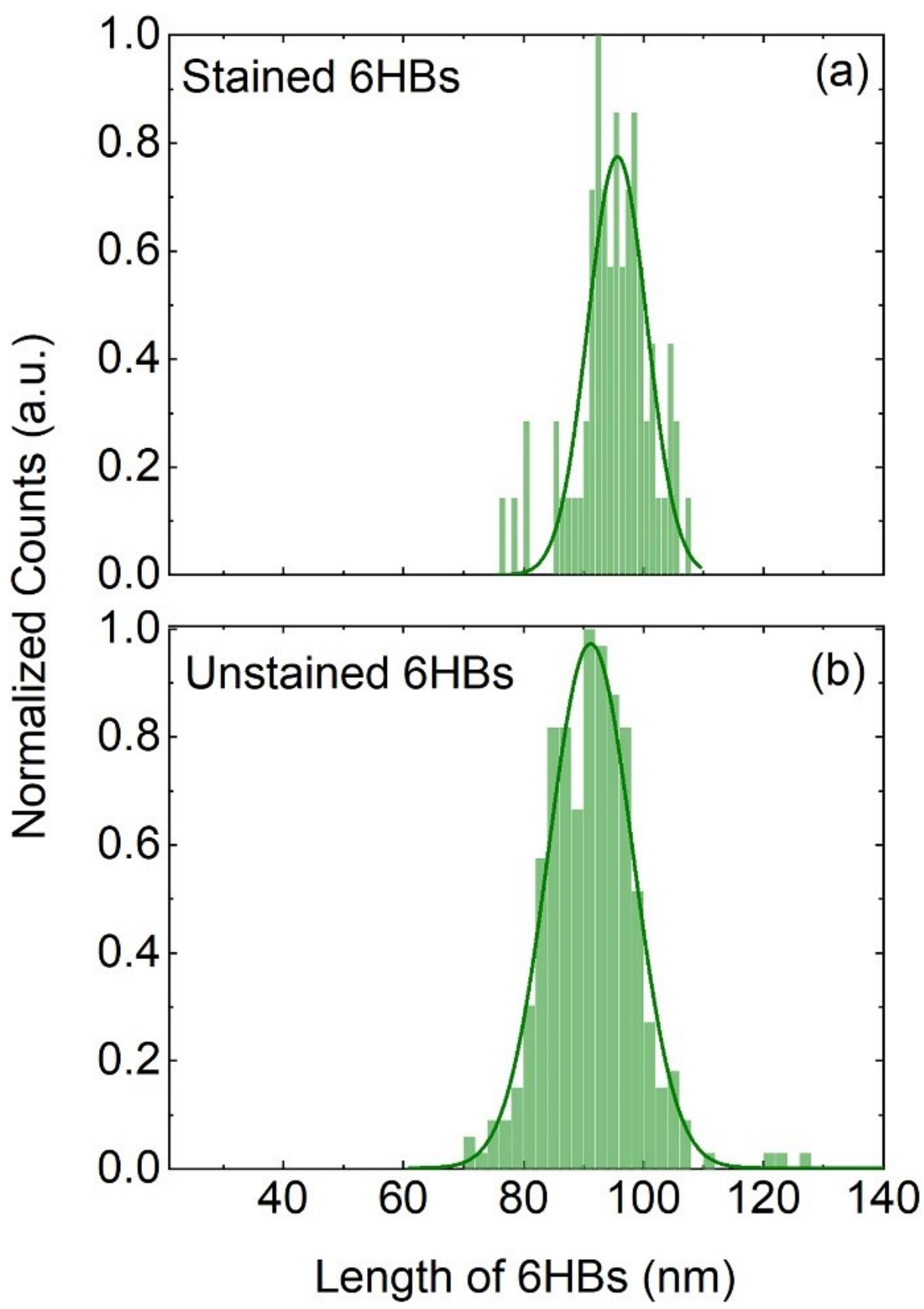

***Figure 6**: Statistical analysis of the impact of staining on the lengths of 6HBs: Histograms of the bundle lengths measured in STEM images of (a) negatively stained and (b) unstained 6HBs. Mean lengths as determined from fits of Gaussian distributions (black lines) to the experimental data are $L_{6HB}$ = 95.7 ± 5.8 nm and $L_{6HB}$ = 91.2 ± 8.3 nm for stained and unstained bundles, respectively. For comparison, the target length predicted from 1702 base pairs per bundle is $L_{6HB,target}$ = 96.4 nm. The result show that the negative staining allows to stabilize the 6HBs with their expected size on the substrate, while unstained, a slight conformational compression is observed.*

# SUPPLEMENTARY INFORMATION

## Design construct of 6-helix bundles (6HBs)

The six-helix bundles (6HBs) used in this study were synthesized as previously described [4,5,29]. In this section, we provide the structural calculations of the 6HB length and connectivity maps between scaffold and staple strands.

As shown in Figure S1, the scaffold has a total length of 1702 nucleotide (nt), which is folded into six parallel DNA helices by assembly with the staple strands. Each helix contains either 283 or 284 base pairs (bp). Given the average rise per base pair in B-form DNA of 0.34 nm/bp, the expected length of a single helix can be estimated. As the helices do not align perfectly in parallel to each other, small variations in lengths are observed across different helices, as illustrated in Figure S1.

Minimum length across a 6HB, $L_{min} = (299-23+1)\times 0.34$ nm = 94.2 nm.

Maximum length across a 6HB, $L_{max} = (306-17+1)\times 0.34$ nm = 98.6 nm.

Mean 6HB length, $L_{mean} = (94.18+98.60)/2$ nm = 96.4 nm.

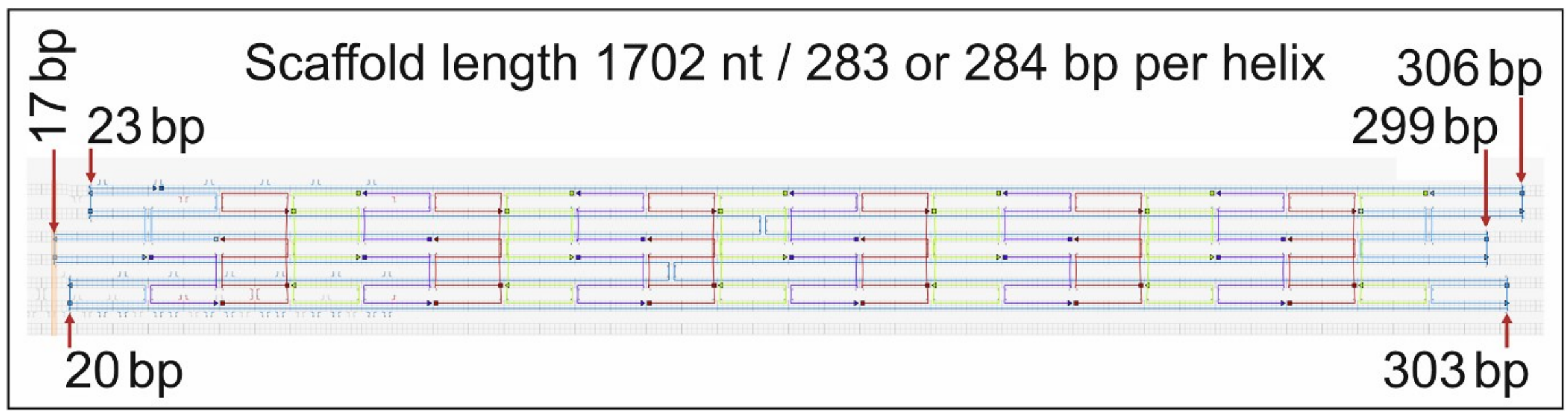

***Figure S1***: *Schematic representation of the scaffold-staple strand connectivity for the 6HBs, illustrating how staple strands direct and stabilize the folding of the scaffold into the designed geometry. A scaffold of 1702 nucleotides (nt) is folded into six helices and stabilized by staple strands. Each helix contains 283 or 284 base pairs (bp). The arrangement of the helices along the scaffold results in minimum and maximum lengths across a 6HB (i.e., the minimum and maximum distances between the front and back ends of the set of helices) of 94.2 nm and 98.6 nm, respectively, which corresponds to a mean length of 96.4 nm.*

## Design construct of 3D molds

The 3-dimentional molds (3D molds) used in this study were synthesized as previously described [6]. In this section, we provide the structural calculations of the 3D mold length and connectivity maps between scaffold and staple strands.

As shown in Figure S2, the scaffold used has a total length of 8064 nucleotide (nt). This correspond to 8064 bases, which were folded into a hollow rectangular box. Each helix contains approximately 126 base pair (bp). Given the average rise per base pair in B-form DNA (~ 0.34 nm/bp), the expected length of 3D mold can be estimated. As the helices forming the box are not align perfectly in parallel, small variations in lengths are observed, as illustrated in Figure S2.

Minimum length across a 3D mold, $L_{min} = (154\text{-}35+1)\times 0.34$ nm $= 40.8$ nm.

Maximum length across a 3D mold, $L_{max} = (162\text{-}27+1)\times 0.34$ nm $= 46.2$ nm.

Mean length across a 3D mold, $L_{mean} = (40{,}80+46{,}24)/2$ nm $= 43.5$ nm.

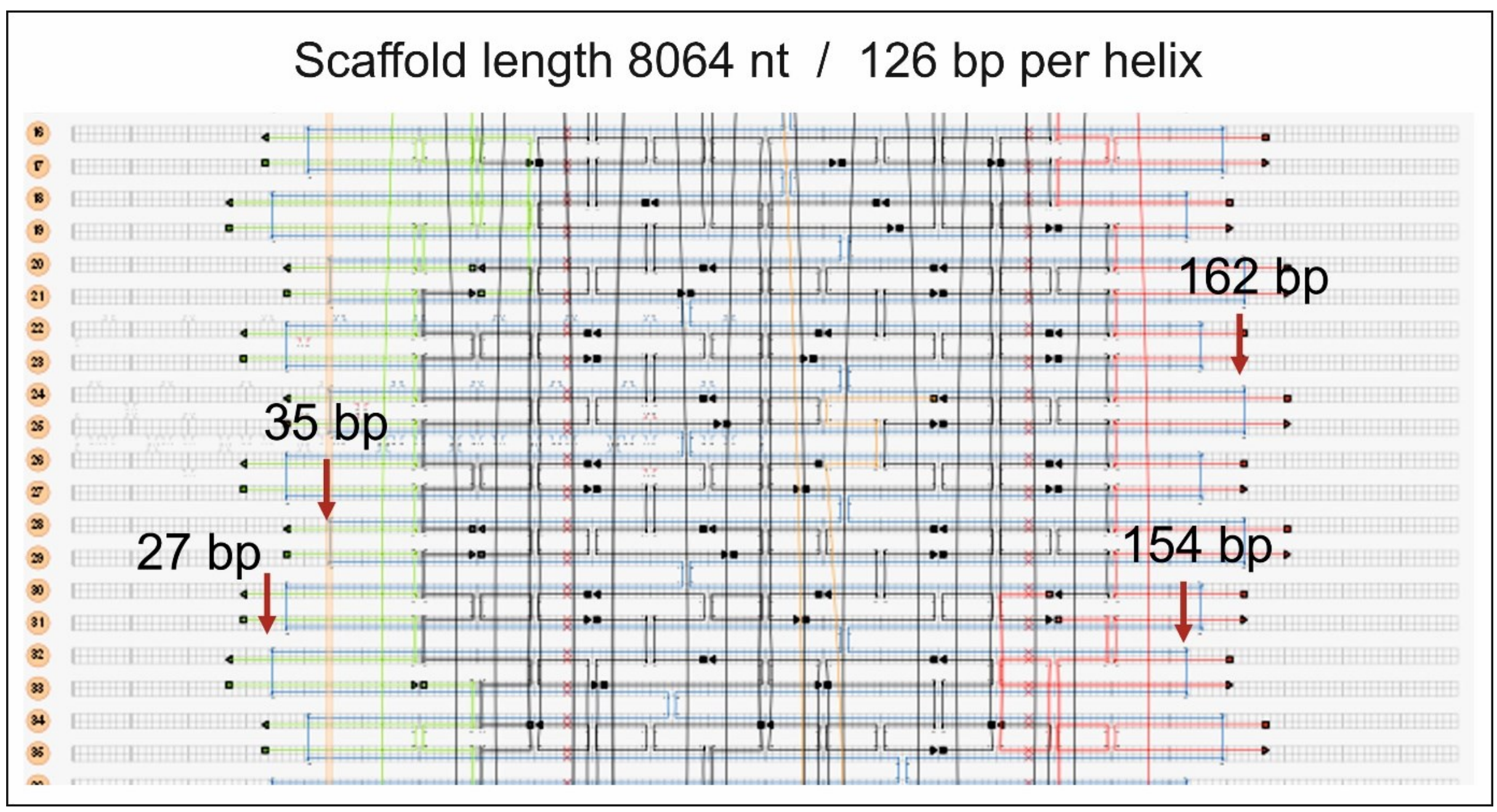

***Figure S2***: *Schematic representation of the scaffold-staple strand connectivity for the 3D molds, illustrating how staple strands direct and stabilize the folding of the scaffold into the designed geometry. A scaffold of 8064 nucleotides (nt) is folded into a hollow rectangular box and stabilized by staple strands. Each helix contains 126 base pairs (bp). The arrangement of the helices along the scaffold results in minimum and maximum lengths across a 3D mold (i.e., the minimum and maximum distances between the front and back ends of the set of helices) of 40.8 nm and 46.2 nm, respectively, which corresponds to a mean length of 43.5 nm.*